\documentclass[twocolumn,showpacs,preprintnumbers,amsmath,amssymb]{revtex4}
\usepackage{graphicx}% Include figure files
\usepackage{dcolumn}% Align table columns on decimal point
\usepackage{bm}% bold math

\begin{document}

\preprint{APS/123-QED}

\title{Comment on ``Turbulent heat transport near critical points: Non-Boussinesq effects''.}% Force line breaks with \\

\author{X. Chavanne}
\affiliation{Institut de Physique du Globe de Paris, G\'eomat\'eriaux et Environnement,\\
4, place Jussieu - Case 89 - 75252 Paris Cedex 05}

%\email{fchilla@ens-lyon.fr}
\author{P.-E. Roche}%
\author{B. Chabaud}%
\author{B. H\'ebral}%
\affiliation{Centre de Recherche sur les Tr\`es Basses Temp\'eratures\\
CNRS, Associ\'e \`a l'Universit\'e Joseph Fourier\\
 25, Av. des Martyrs, BP 166, 38042 Grenoble Cedex 9}
\author{F. Chill\`a}
\author{B. Castaing}%
\affiliation{Ecole Normale Sup\'erieure de Lyon,\\
46, All\'ee d'Italie, 69364
Lyon Cedex 7, France}%

\date{9/march/2006}% It is always \today, today,

\begin{abstract}
In a recent preprint ( cond-mat/0601398 ), D. Funfschilling and G. Ahlers describe a new effect, that they interpret as non-Boussinesq, in a convection cell working with ethane, near its critical point. They argue that such an effect could have spoiled the Chavanne {\it et al.} (Phys. Rev. Lett. {\bf 79} 3648, 1997) results, and not the Niemela {\it et al.} (Nature, {\bf 404}, 837, 2000) ones, which would explain the differences between these two experiments. We show that:\\
-i)Restricting the Chavanne's data to situations as far from the critical point than the Niemela's one, the same discrepancy remains.\\
-ii)The helium data of Chavanne show no indication of the effect observed by  D. Funfschilling and G. Ahlers.
\end{abstract}

\maketitle

The question of whether or not the fully turbulent Kraichnan regime \cite{Kraichnan,Siggia} has been observed in Rayleigh-B\'enard (RB) convection is a subject of vigorous controversy in the past decade. Chavanne {\it et al.}\cite{Chavanne1,Chavanne2} observed a rapid increase of the Nusselt number $Nu$ versus the Rayleigh number $Ra$, and interpreted it as the transition toward this ultimate regime. The difference between the observed logarithmic slope of this dependence and the predicted $1/2$ exponent was attributed to the logarithmic correction proposed by Kraichnan himself \cite{Kraichnan,Siggia}. Roche {\it et al.} \cite{Roche} showed that rough boundaries fix the correction as expected and induce a $1/2$ exponent behaviour. Later, Niemela and Sreenivasan \cite{Niemela2} reported the same kind of behaviour.

 \begin{figure}[!h]

   \begin{center}
   \includegraphics[width= 8cm]{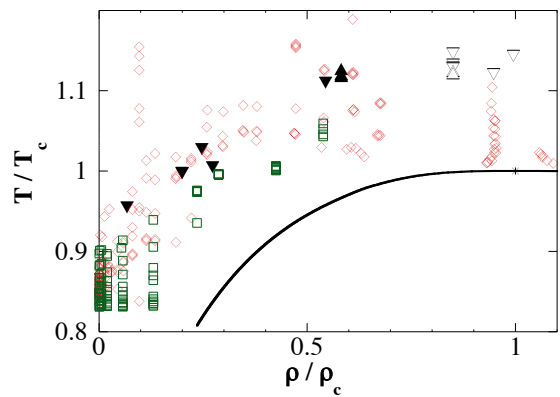}
   \end{center}

\caption{Phase diagram of $^4$He in reduced units ($T_c=5.1953$ K, $\rho_c=69.641$ kgm$^{-3}$) showing the location of various measurements of $Nu$ versus $Ra$. Solid line: coexistence curve. Asterisc: critical point. Open diamonds: Niemela {\it et al.} \cite{Niemela1}. Open squares: Wu {\it et al.} \cite{Wu1,Wu2}. Triangles up and down: Chavanne {\it et al.} \cite{Chavanne2}; triangles up: $\beta\Delta T<0.1$, triangles down: $\beta\Delta T>0.1$, open: near the critical isochore, solid: far from the critical isochore.}

\label{Tdens}

\end{figure}

However Wu {\it et al.} \cite{Wu1,Wu2} and Niemela {\it et al.} \cite{Niemela1} explored the same range of $Ra$, with the same working fluid (helium) without observing the same increase in $Nu$. Several propositions have been made, but there exist no clear consensus about the differences between the cells which could explain such a discrepancy.

In a recent work, D. Funfschilling and G. Ahlers (FA) \cite{FA} observe a new effect, that they interpret as spurious (non-Boussinesq), in their convection cell working with ethane, near its critical point. They remark that most of the working points of the works observing the Kraichnan like behaviour \cite{Chavanne1,Chavanne2,Niemela2,Roche} are nearer the critical point and/or the gas-liquid coexistence curve of helium than those not observing it \cite{Wu1,Wu2,Niemela1}. They thus propose that this spurious effect explains the difference quoted above, which would mean that the Kraichnan regime has not yet been experimentally observed.

In their figure 4, they indicate the position of the different working points in the phase space. It can be seen that some of the points of Chavanne {\it et al.} are within the points of Niemela {\it et al.}. In our figure \ref{Tdens} we reproduce only these points (triangles up and down). The temperature $T$ is $T_m=(T_b+T_t)/2$, where $T_t$ (resp. $T_b$) is the top (resp. bottom) plate temperature. $\rho$ is the fluid average density, $T_c$ and $\rho_c$ are the critical temperature and density.  We also show the Niemela {\it et al.} \cite{Niemela1} points (diamonds), recently provided to us by J.J. Niemela, and the Wu {\it et al.} \cite{Wu1,Wu2} points (squares). We note that some of the points of Niemela {\it et al.}, not shown on the FA figure 4, are very near the critical point, however not showing the Kraichnan like increase in $Nu$ \cite{Niemela1}.

 \begin{figure}[!h]

   \begin{center}
   \includegraphics[width= 8cm]{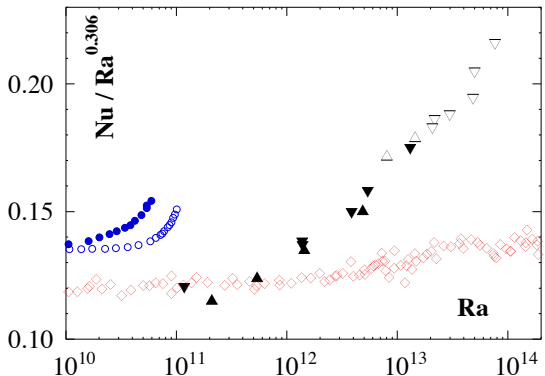}
   \end{center}

\caption{The compensated Nusselt versus the Rayleigh number. Circles: FA results as presented in their figure 5 (open: isothermal, solid: isobaric with $\Delta T=1$ K, $P/P_c=0.920$). Other symbols are as in figure \ref{Tdens}.}

\label{NuRaloin}

\end{figure}

In figure \ref{Tdens}, and in the following one \ref{NuRaloin} we do several distinctions between the Chavanne's points we kept. Points with $\beta\Delta T<0.1$, where $\beta$ is the isobaric expansion coefficient and $\Delta T=T_b-T_t$, are plotted as triangles up. We can consider for these points that physical properties are reasonably uniform within the cell. Points with $\beta\Delta T>0.1$ are plotted as triangles down. To eliminate any doubt, we also distinguish densities far from the critical one (solid symbols) from those near it (open symbols).

Following FA (\cite{FA} figure 3), we plot in our figure \ref{NuRaloin} $Nu_{comp}=Nu/Ra^{0.306}$ {\it versus} $Ra$ for the Chavanne's points we kept. The symbols are the same than in figure  \ref{Tdens}. Also shown are the FA results (circles) as presented in their figure 5 (open: isothermal, solid: isobaric with $\Delta T=1$ K), and the Niemela's points \cite{Niemela1}. The discrepancy remains for these points which have the same location in the phase space. The Kraichnan like increase for the Chavanne's points is obvious. We can easily conclude that the discrepancy between references \cite{Chavanne1,Chavanne2} and \cite{Niemela1} has nothing to do with the location in the phase space.

 \begin{figure}[!h]

   \begin{center}
   \includegraphics[width= 8cm]{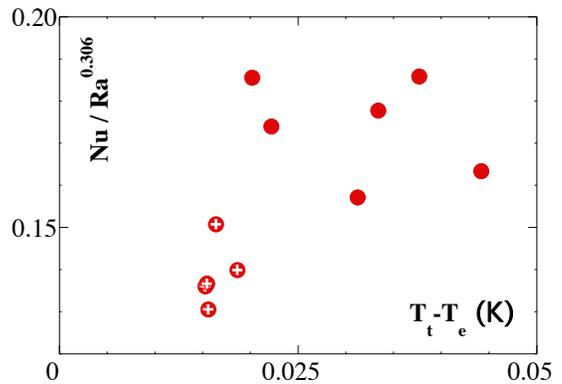}
   \end{center}

\caption{Compensated Nusselt versus the difference between the top plate temperature and the condensation temperature at the considered pressure. All the points correspond to the same density: $37.9$ kgm$^{-3}$. The circles with cross correspond to the same pressure within $\pm0.2$\% ($P= 1.955$ bar, $P/P_c=0.86$)}

\label{isodensT}

\end{figure}

Another question is to know if whether or not the characteristics of the effect observed by Funfschilling and Ahlers can be found in the data of Chavanne {\it et al.}. Funfschilling and Ahlers have shown that, in their observations, the compensated Nusselt $Nu_{comp}=Nu/Ra^{0.306}$ only depends on the difference between $T_t$ (not $T_m$) and the equilibrium temperature $T_e$ at the same pressure (this corresponding to $T_m$ and $\rho$). 
 
\begin{figure}[!h]

   \begin{center}
   \includegraphics[width= 8cm]{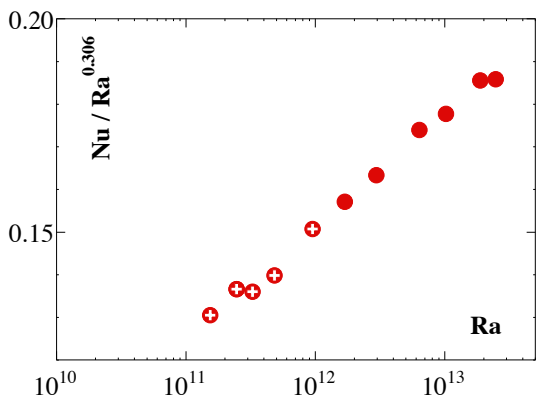}
   \end{center}

\caption{Same points as figure \ref{isodensT} plotted versus $Ra$. }

\label{isodensRa}

\end{figure}

There are no constant pressure series of measurements in Chavanne's data \cite{Chavanne1,Chavanne2}. The reason is that ensuring a constant pressure ask for letting the filling capillary open up to a pressure gauge or ballast. On some points of the capillary, the temperature can go down to the condensation temperature $T_e$. Together with the important compressibility near the critical point, it can cause large transfers of mass and heat in this capillary. However
several series of measurements in  \cite{Chavanne1,Chavanne2} have been made at constant density $\rho$. On the figure \ref{isodensT} we plot the compensated Nusselt versus $(T_t-T_e)$ for the Chavanne's points corresponding to $\rho=37.9$ kgm$^{-3}$. The strongly different values do not merge on a curve. On the opposite, they nicely merge if plotted versus $Ra$ (figure \ref{isodensRa}). We draw the attention of the reader on the points labeled with a cross in the circle. Their corresponding pressure is the same within $\pm0.2$\%, which is not as precise as the $\pm3\,10^{-5}$ of Funfschilling and Ahlers, but rather constant anyway.

To conclude, the simple examination of the published data of Chavanne {\it et al.} shows that the hypothesis raised by Funfschilling and Ahlers \cite{FA}, that the influence of the critical point could explain the apparent discrepancy between references \cite{Chavanne1,Chavanne2} and \cite{Niemela1}, this hypothesis cannot hold. Moreover, as far as the effect observed by them \cite{FA} is well characterized, we found no trace of it in the helium data of \cite{Chavanne1,Chavanne2}. We see here no reason to reconsider the original interpretation for the increase of $Nu$ observed by \cite{Chavanne1,Chavanne2,Roche} as a transition toward the Kraichnan regime, even if the exact conditions for such a transition have yet to be elucidated.

\begin{acknowledgments}
We are grateful to J.J. Niemela for providing us the numerical values corresponding to the data in Ref. \cite{Niemela1}. We also thank G. Ahlers for supplementary details on the cell and on ethane properties.

\end{acknowledgments}

\end{document}